\documentstyle[12pt]{article}

\begin{document}
\begin{flushright}
DCE-APU 97-01 \\
AUE 97/01 \\
DPNU-97-24 \\
hep-ph/9705402 \\
May 1997\vspace{1cm}
\end{flushright}

\begin{center}
{\bf\large New Possibility of Solving the Problem of Lifetime 
Ratio $\tau ({\sl \Lambda}_b ) / \tau (B_d)$} 
\vspace{1cm}

 Toshiaki Ito\footnote[2]{toshiaki@eken.phys.nagoya-u.ac.jp}, 
~Masahisa Matsuda$^*$\footnote[3]{mmatsuda@auecc.aichi-edu.ac.jp} 
and ~Yoshimitsu Matsui$^{**}$\footnote[4]{matsui@eken.phys.nagoya-u.ac.jp}
\vspace{1cm}

\sl{Department of Childhood Education, Aichi Prefectural University,} 
{\it Nagoya 467, Japan}\\
$^*$\sl{Department of Physics and Astronomy, \\
Aichi University of Education,} {\it Kariya 448, Japan}\\
$^{**}$\sl{Department of Physics, Nagoya University,} 
{\it Nagoya 464-01, Japan}

\end{center}

\begin{abstract}

We discuss the problem of the large discrepancy between the observed lifetime 
ratio of  ${\sl \Lambda}_b$ to $B_d$ and the theoretical prediction obtained 
by the heavy quark effective theory. A new possibility of solving this 
problem is proposed from the viewpoint of 
operator product expansion and the lifetime ratio of 
${\sl \Omega}_b$ to $B_d$ is predicted.
\end{abstract}
\vspace{0.5cm}


The heavy quark effective theory (HQET) is successful in explaining various 
nature of the hadrons containing a heavy quark. The HQET has been extensively
applied to the meson system containing a heavy quark and has brought about 
many remarkable results. 
The HQET has been also applied to the system of  baryons containing a heavy 
quark and has led to some interesting results.
However, when applying the HQET to the heavy baryon systems, 
two problems are encountered.
One is that the experimental value  of the mass difference between  
${\sl \Sigma}_b$ and ${\sl \Sigma}_b^{*}$ is quite large in
comparison with the predicted value from the mass relation in the HQET as
\begin{eqnarray}
   m_{{\sl \Sigma}_{b^*}}-m_{{\sl \Sigma}_b} & = & 56 \pm 13 \, \mbox{MeV} 
                            \quad (Exp.) \,\cite{DELPHI1} , \nonumber\\
                        & = & 15.8 \pm 3.3 \, \mbox{MeV} 
                        \quad (Theory)\,\cite{Jenkins1} .
\end{eqnarray}                             
The other one is that 
the experimental lifetime ratio of ${\sl \Lambda}_b$ to $B_d$ \cite{LEP1}
\begin{equation}
  \frac{\tau\left({\sl \Lambda}_b \right)}{\tau \left( B_d \right)} = 
   0.78 \pm 0.06 
  \label{eqn:ratioblexp}
\end{equation}
is quite small compared with the theoretical 
predictions\cite{BBSUV1}--\cite{Colan1}.

In this paper, we concentrate our attention to the latter problem and 
propose a possible solution to this problem from the viewpoint of operator 
product expansion(OPE). Moreover, we point out the importance of measuring 
the lifetime of ${\sl \Omega}_b$ to select out the models.
\vspace{1cm}


The inclusive decay width of a hadron $H_b$ containing a bottom quark can be
written as \cite{NS1}
\begin{equation}
   \Gamma(H_b\to X_f) = {1\over m_{H_b}}\,\mbox{Im}\,
   \langle H_b|\,\hat{T}\,|H_b\rangle,\label{eqn:gamma}
\end{equation}
where the transition operator $\hat{T}$ is given by 
\begin{equation}
   \hat{T} = i\!\int d^4 x\,\mbox{T}\{\,
   {\cal L}_W (x) \, {\cal L}_W (0)\,\} \, . \label{eqn:TransOP}
\end{equation}
The effective Lagrangian ${\cal L}_W$  for the weak decay $ H_b \to X_f$ is 
\begin{eqnarray}
   {\cal L}_W &=& - {4 G_F\over\sqrt{2}}\,V_{cb}\,
    \bigg\{ c_1(m_b)\,\Big[
    \bar d'_L\gamma_\mu u_L\,\bar c_L\gamma^\mu b_L +
    \bar s'_L\gamma_\mu c_L\,\bar c_L\gamma^\mu b_L \Big] \nonumber\\
   &&\phantom{ - {4 G_F\over\sqrt{2}}\,V_{cb}\, }
    \mbox{}+ c_2(m_b)\,\Big[
    \bar c_L\gamma_\mu u_L\,\bar d'_L\gamma^\mu b_L +
    \bar c_L\gamma_\mu c_L\,\bar s'_L\gamma^\mu b_L \Big] \nonumber\\
   &&\phantom{ - {4 G_F\over\sqrt{2}}\,V_{cb}\, }
    \mbox{}+ \sum_{\ell=e,\mu,\tau}
    \bar\ell_L\gamma_\mu\nu_\ell\,\bar c_L\gamma^\mu b_L
    \bigg\} + \mbox{h.c.} \,, \label{eqn:EffLag}
\end{eqnarray}
where $q_L$ denotes a left-handed quark field and  
$d^\prime$ and $s^\prime$ stand for the weak eigenstates. 
We neglect the CKM suppressed transitions $b \to u$, because this effect 
is negligibly small in the present analyses.
Up to the leading order, the combinations 
$c_{\pm} = c_1 \pm c_2 $ are given by 
\begin{equation}
   c_\pm(m_b) = \left( {\alpha_s(m_W)\over\alpha_s(m_b)}
   \right)^{a_\pm} \,,\qquad
   a_- = -2 a_+ = - {12\over 33 - 2 n_f} \,.\label{eqn:cpm}
\end{equation}

In order to pursue the calculation of the inclusive decay width in 
eq.(\ref{eqn:gamma}), we must evaluate the matrix element of the nonlocal 
operator $\hat{T}$ in eq.(\ref{eqn:TransOP}). 
Although we don't know how to evaluate such the matrix element of the
nonlocal operator directly, it is efficient for us to use the method of the OPE
for the $\hat{T}$. This is because the energy release is quite large 
compared with the QCD scale in the bottom quark decay.
Using the OPE, the total decay width of a hadron $H_b$ 
can be written in the form \cite{BS1}
\begin{eqnarray}
   \Gamma(H_b\to X_f) = {G_F^2 m_b^5\over 192\pi^3}\,
   {1\over 2 m_{H_b}}\,\Bigg\{
   c_3(f)\,\langle H_b|\,\bar b b\,|H_b\rangle 
   + c_5(f)\,\frac{\langle H_b|\,\bar b\,g_s\sigma_{\mu\nu} G^{\mu\nu}
   b\,|H_b\rangle}{\Delta^2} \nonumber \\
   + \sum_i c_6^i(f)\frac{\langle H_b|\,(\bar b\,\Gamma_i q)\,(\bar q
   \Gamma_i b)
   \,|H_b\rangle}{\Delta^3} + \dots \Bigg\} \,,\label{eqn:gener}
\end{eqnarray}
where $c_n(f)$ are dimensionless coefficient functions depending on 
the quantum numbers of the final state and including the renormalization 
group and phase factors,  $\Gamma_i$'s denote the 
combinations of 
$\gamma$-matrices and $\Delta$ is an expansion parameter with the
mass dimension, which should be much larger than the scale of
$\Lambda_{QCD}$ to justify the OPE.

  The Dirac spinor $b(x)$ in 
eq.(\ref{eqn:gener}), which is an operator of QCD, can be expressed as
\begin{equation}
    b\,(x) = e^{-im_b v \cdot x} \left\{ h_b(x) + \chi_b(x) \,\right\} \, ,
\end{equation}
where $h_b(x)$ and $\chi_b(x)$ are the large and small component
of the spinor $b(x)$ respectively and $v$ stands for the velocity of 
the hadron containing a bottom quark\cite{Neubert1}. 
The $\chi_b(x) $ can be written 
in terms of $h_b(x)$ by using the equation of motion. Therefore the
first and second matrix elements in eq.(\ref{eqn:gener}) are
expanded as \cite{BBSUV1,NS1} 
\begin{eqnarray}
   \frac{1}{2m_{H_b}}\langle H_b|\,\bar b b\,|H_b\rangle & = & 1 
   - \frac{\mu_{\pi}^2(H_b) 
   - \mu_G^2(H_b)}{4m_b^2} + {\cal O}\left( \frac{1}{m_b^3} \right) 
   \,, \label{eqn:HQET1}\\
   \frac{1}{2m_{H_b}}\langle H_b|\,\bar b\,g_s\sigma_{\mu\nu} G^{\mu\nu} b\,
   |H_b\rangle & = & 2 \mu_G^2(H_b) +
   {\cal O} \left( \frac{1}{m_b} \right) \,, \label{eqn:HQET2}
\end{eqnarray}
with $\mu_\pi^2 (H_b)$ and $\mu_G^2 (H_b)$ defined by
\begin{eqnarray}
   \mu_\pi^2 (H_b)  & \equiv & - \frac{1}{2m_{H_b}} \langle H_b|\,
   \bar{h}_b\,\left(iD_{\bot}\right)^2 h_b\,|H_b\rangle \,, \\
   \mu_G^2 (H_b) & \equiv & \frac{1}{4m_{H_b}} 
   \langle H_b|\,\bar h_b\,g_s\sigma_{\mu\nu} G^{\mu\nu} h_b\,|H_b\rangle \,,
\end{eqnarray}
where $D_{\bot}^{\mu} = \partial^{\mu} - v^{\mu} (v \cdot D)$ and $D^{\mu}$ 
is the covariant derivative of QCD. 
The parameters $\mu_\pi^2 (H_b)$ and $\mu_G^2 (H_b)$ 
represent the matrix elements of the kinetic energy operator and 
the chromo-magnetic one, respectively. 
These parameters can be estimated by using the mass spectra of 
heavy hadron states and are of ${\cal O}( \Lambda_{QCD}^2 )$.

The third matrix element in eq.(\ref{eqn:gener}) can be parameterized in
the model-independent way\cite{NS1}. For the meson matrix elements of 
local four-quark operators, we use the same parameters $B_i$ and 
$\varepsilon_i$ as Ref.\cite{NS1} such that 
\begin{eqnarray}\label{eqn:mesons}
   \frac{1}{2 m_{B_q}}\,\langle B_q|\,O_{V-A}^q\,|B_q\rangle
   &\equiv& \frac{f_{B_q}^2 m_{B_q}}{8}\,B_1 \,, \nonumber\\
   \frac{1}{2 m_{B_q}}\,\langle B_q|\,O_{S-P}^q\,|B_q\rangle
   &\equiv& \frac{f_{B_q}^2 m_{B_q}}{8}\,B_2 \,, \nonumber\\
   \frac{1}{2 m_{B_q}}\,\langle B_q|\,T_{V-A}^q\,|B_q\rangle
   &\equiv& \frac{f_{B_q}^2 m_{B_q}}{8}\,\varepsilon_1 \,,
    \nonumber\\
   \frac{1}{2 m_{B_q}}\,\langle B_q|\,T_{S-P}^q\,|B_q\rangle
   &\equiv& \frac{f_{B_q}^2 m_{B_q}}{8}\,\varepsilon_2 \,,
\end{eqnarray}
where $f_{B_q}$ is the decay constant of $B_q$ meson. The local four-quark
operators defined by
\begin{eqnarray}\label{eqn:4qops}
   O_{V-A}^q &=& \bar b_L\gamma_\mu q_L\,\bar q_L\gamma^\mu b_L
    \,, \nonumber\\
   O_{S-P}^q &=& \bar b_R\,q_L\,\bar q_L\,b_R \,, \nonumber\\
   T_{V-A}^q &=& \bar b_L\gamma_\mu t_a q_L\,\bar q_L\gamma^\mu
    t_a b_L \,, \nonumber\\
   T_{S-P}^q &=& \bar b_R\,t_a q_L\,\bar q_L\,t_a b_R \,,
\end{eqnarray}
where $t_a = \lambda_a / 2 $ are the generators of color SU(3). 
For the baryon matrix elements of local four-quark operators, we use 
the same parameters as Ref.\cite{NS1,Cheng1} such that 
\begin{eqnarray}\label{eqn:Lb}
   \langle\Lambda_b|\,\widetilde O_{V-A}^q\,|\Lambda_b\rangle &\equiv&
   - \widetilde B_{{\sl \Lambda}_b}\,\langle\Lambda_b|\,O_{V-A}^q\,
   |\Lambda_b\rangle \,,\nonumber\\
   \frac{1}{2 m_{\Lambda_b}}\,\langle\Lambda_b|\,O_{V-A}^q\,
   |\Lambda_b\rangle &\equiv& - \frac{f_{B_q}^2 m_{B_q}}{48}\,
   r_{{\sl \Lambda}_b} \,, \nonumber \\
   \langle\Omega_b|\,\widetilde O_{V-A}^q\,|\Omega_b\rangle &\equiv&
   - \widetilde B_{{\sl \Omega}_b}\,\langle\Omega_b|\,O_{V-A}^q\,
   |\Omega_b\rangle \,,\nonumber\\
   \frac{1}{2 m_{\Omega_b}}\,\langle\Omega_b|\,O_{V-A}^q\,
   |\Omega_b\rangle &\equiv& - \frac{f_{B_q}^2 m_{B_q}}{8}\,
   r_{{\sl \Omega}_b} \,,
\end{eqnarray}
where the local four-quark operator $\widetilde O_{V-A}^q$ 
is defined by
\begin{equation}
   \widetilde O_{V-A}^q = \bar b_L^i\gamma_\mu q_L^j\,
    \bar q_L^j\gamma^\mu b_L^i \,.
\end{equation}

By using these parameters, the $1/\Delta^3$-term of each bottom hadron 
can be written as follows.
\begin{eqnarray}\label{eqn:dim6}
 \lefteqn{
   \frac{1}{2 m_B} \sum_i c_6^i(f)\langle B^-|\,(\bar b\,\Gamma_i q)\,
   (\bar q \Gamma_i b) \,|B^-\rangle } \nonumber \\
   &&= \eta \,(1-z)^2\,\left\{
    (2 c_+^2 - c_-^2)\,B_1 + 3 (c_+^2 + c_-^2)\,\varepsilon_1 \right\}
    \,,  \nonumber\\
 \lefteqn{
   \frac{1}{2 m_B} \sum_i c_6^i(f)\langle B_d|\,(\bar b\,\Gamma_i q)\,(\bar q
   \Gamma_i b) \,|B_d\rangle} \nonumber \\
    &&= - \eta \,(1-z)^2\,
    \Bigg\{ \frac{1}{3}\,(2 c_+ - c_-)^2\,\left[
    \left( 1 + \frac{z}{2} \right)\,B_1 - (1+2z)\,B_2 \right]
    \nonumber\\
   &&\quad\mbox{}+ \frac{1}{2}\,(c_+ + c_-)^2\,
    \bigg[ \left( 1 + \frac{z}{2} \right)\,\varepsilon_1
    - (1+2z)\,\varepsilon_2 \bigg] \Bigg\} \nonumber\\
   &&\quad\mbox{}- \eta\,\sqrt{1-4z}\,
    \frac{|V_{cd}|^2}{|V_{ud}|^2}\,\Bigg\{ {1\over 3}\,(2 c_+ - c_-)^2\,
    \left[(1-z)\,B_1 - (1+2z)\,B_2 \right] \nonumber\\
   &&\quad\mbox{}+ \frac{1}{2}\,(c_+ + c_-)^2\,\left[
    (1-z)\,\varepsilon_1 - (1+2z)\,\varepsilon_2 \right] \Bigg\} \,,\nonumber\\
 \lefteqn{
   \frac{1}{2 m_{B_s}} \sum_i c_6^i(f)\langle B_s|\,(\bar b\,\Gamma_i q)\,
   (\bar q \Gamma_i b) \,|B_s\rangle} \nonumber \\
    &&= - \eta' \,(1-z)^2\,\frac{|V_{us}|^2}{|V_{cs}|^2}\,
    \Bigg\{ \frac{1}{3}\,(2 c_+ - c_-)^2\,\left[
    \left( 1 + \frac{z}{2} \right)\,B_1 - (1+2z)\,B_2 \right]
    \nonumber\\
   &&\quad\mbox{}+ {1\over 2}\,(c_+ + c_-)^2\,\left[
    \left( 1 + \frac{z}{2} \right)\,\varepsilon_1
    - (1+2z)\,\varepsilon_2 \right] \Bigg\} \nonumber\\
   &&\quad\mbox{}- \eta'\,\sqrt{1-4z}\,
    \Bigg\{ {1\over 3}\,(2 c_+ - c_-)^2\,\left[
    (1-z)\,B_1 - (1+2z)\,B_2 \right] \nonumber\\
   &&\quad\mbox{}+ \frac{1}{2}\,(c_+ + c_-)^2\,\left[
    (1-z)\,\varepsilon_1 - (1+2z)\,\varepsilon_2 \right] \Bigg\} \,,
    \nonumber\\
 \lefteqn{\frac{1}{2 m_{\Lambda_b}}\sum_i c_6^i(f)\langle\Lambda_b|\,
   (\bar b\,\Gamma_i q)\,(\bar q \Gamma_i b)\,|\Lambda_b\rangle} \nonumber\\
   && = \eta\, \frac{r_{{\sl \Lambda}_b}}{16}\,\Bigg\{
    8 (1-z)^2\,\left[ (c_-^2 - c_+^2)
    + (c_-^2 + c_+^2)\,\widetilde B_{{\sl \Lambda}_b} \right] \nonumber\\
   &&\mbox{}- \left[ (1-z)^2 (1+z)\,
    + \sqrt{1-4 z}\,\frac{|V_{cd}|^2}{|V_{ud}|^2} \, \right] \nonumber\\
   &&\quad\times \left[ (c_- - c_+)(5 c_+ - c_-)
    + (c_- + c_+)^2\,\widetilde B_{{\sl \Lambda}_b} \right] \Bigg\} \,,
    \nonumber \\
 \lefteqn{\frac{1}{2 m_{\Omega_b}}\sum_i c_6^i(f)\langle\Omega_b|\,
   (\bar b\,\Gamma_i q)\,(\bar q \Gamma_i b)\,|\Omega_b\rangle} \nonumber\\
   && = - \eta'\, \frac{r_{{\sl \Omega}_b}}{24}
     \sqrt{1-4 z}\, \left[ (c_- - c_+)(5 c_+ - c_-)
    + (c_- + c_+)^2\,\widetilde B_{{\sl \Omega}_b} \right] \,,
\end{eqnarray}
where $z \equiv m_c^2 / m_b^2$ (we set z equal to 0.083 \cite{NS1}) and 
$\eta$ and $\eta'$ are defined by
\begin{eqnarray}
\eta & \equiv & 16 \pi^2 f_B m_B |V_{cb}|^2 |V_{ud}|^2 \,, \nonumber \\
\eta' & \equiv & 16 \pi^2 f_{B_s} m_{B_s} |V_{cb}|^2 |V_{cs}|^2 \,,
\end{eqnarray}
respectively. We set both of $f_{B}$ and $f_{B_s}$ equal to 
$0.18\,\mbox{GeV}$ \cite{Sach1}.


By using these model independent parameters $B_i$, $\varepsilon_i$,
$\widetilde B_{{\sl \Lambda}_b}$ and $r_{{\sl \Lambda}_b}$,  
the lifetime ratio of ${\sl \Lambda}_b$ to $B_d$
can be written as
\begin{eqnarray}
   \frac{\tau \left({\sl \Lambda}_b\right)}{\tau \left(B_d\right)} &=& 1
   + \frac{\mu_\pi^2 \left({\sl \Lambda}_b\right) 
   - \mu_\pi^2 \left(B_d\right)}{4 m_b^2} 
   + \left(\frac{1}{4} + c_M
   \frac{m_b^2}{\Delta^2}\right)
   \frac{\mu_G^2 \left(B_d\right) 
   - \mu_G^2 \left({\sl \Lambda}_b\right)}{m_b^2} \nonumber \\
   &&+ \frac{1}{\Delta^3}\left\{k_1 B_1 + k_2 B_2 + k_3 \varepsilon_1 
   + k_4 \varepsilon_2  + \left( k_5+ 
     k_6 \widetilde B_{{\sl \Lambda}_b} \right) r_{{\sl \Lambda}_b} \right\} 
     \,,  \label{eqn:ratiobl}
\end{eqnarray}
where $c_M$ and $k_i$'s are the coefficients including the renormalization 
group factors $c_{\pm}$ and phase factor $z$. By using the same definition 
of $A_i$ and $z_i$ as in Ref.\cite{BUV1}, $c_M$ is given by
\begin{eqnarray}
  c_M &\equiv& \frac{2c_5(f)}{c_3(f)} \\
      &\cong& - \frac{2 \left(N_c A_0 z_1 + 4 N_c A_2 z_2 + 2 z_1 \right)}
      {N_c A_0 z_0 + 2 z_0} \,.
      \label{eqn:cm}
\end{eqnarray}
At the heavy quark limit 
$m_b \rightarrow \infty$, the lifetime ratio (\ref{eqn:ratiobl}) 
approaches unity. However, the recent experiment (\ref{eqn:ratioblexp}) shows 
that the large discrepancy exists between the $B_d$ and ${\sl \Lambda}_b$ 
lifetimes.  Recently several literatures \cite{BBSUV1}--\cite{Colan1} have 
been devoted to study this problem. To seek a solution that can reconcile 
this discrepancy, the contributions of the ${\cal O} \left( 1/\Delta^3 \right)$
term in eq.(\ref{eqn:gener}) have been estimated under the condition that 
the expansion parameter $\Delta$ is equal to $m_b$. 
This condition might be natural because the physical scales in the system are 
only $\Lambda_{QCD}$ and $m_b$. 
The parameters $\mu^2_\pi(H_b)$ and $\mu^2_G(H_b)$ in eq.(\ref{eqn:ratiobl}) 
can be estimated from the mass formulae of the hadrons 
containing a bottom quark as follows
\begin{eqnarray}
   \mu_\pi^2({\sl \Lambda}_b) - \mu_\pi^2(B_d) &=&
    - (0.01\pm 0.03)~(\mbox{GeV})^2 \,, \nonumber\\
   \mu_G^2(B_d) &=& {3\over 4}\,(m_{B^*}^2 - m_B^2)
    \simeq 0.36~(\mbox{GeV})^2 \,, \nonumber\\
   \mu_G^2({\sl \Lambda}_b) &=& 0 \,.\label{eqn:muvals}
\end{eqnarray}
The estimation of the parameters $B_i$, $\varepsilon_i$, 
$\widetilde B_{{\sl \Lambda}_b}$ and $r_{{\sl \Lambda}_b}$ has only
been carried out in the model-dependent ways. 
The parameter $\widetilde B_{{\sl \Lambda}_b}$ is equal to 1 in valence 
quark approximation. Because the matrix elements of $O_{V-A}^q$ and 
$\widetilde O_{V-A}^q$ differ only by a sign in this approximation, 
since the color wave function for a baryon is totally antisymmetric.
The parameters $B_i$ and $\varepsilon_i$ have been estimated
by using QCD sum rules\cite{BLLS1}. The parameter $r_{{\sl \Lambda}_b}$
have been estimated by using both QCD sum rules\cite{Colan1} and 
non-relativistic quark model\cite{Rosner1,Cheng1}. These analyses give 
the result in the ratio
\begin{equation}
   \frac{\tau\left({\sl \Lambda}_b \right)}{\tau \left( B_d \right)} 
   \geq 0.94.  
   \label{eqn:ratcr}
\end{equation}
Thus the problem remains unsolved at this stage.

One possibility of solving this problem was proposed in Ref.\cite{AMPR1}. 
This proposal contains the insistence that the mass $m_b$ in the factor 
$m_b^5$ in eq.(\ref{eqn:gener}) should be replaced by 
the mass of the parent hadron $m_{B_d}$ or $m_{{\sl \Lambda}_b}$.
If this insistence is correct, the lifetime ratio of $B_d$ to 
${\sl \Lambda}_b$ is almost determined by the ratio of the 
decaying hadron masses;  
\begin{equation}
  \frac{\tau\left({\sl \Lambda}_b \right)}{\tau \left( B_d \right)} \sim 
  \left(\frac{m_B}{m_{{\sl \Lambda}_b}} \right)^5 = 0.73 \pm 0.01\,.
\end{equation}
When this proposal is applied to charmed hadrons lifetime, the results
are in good agreement with the experiment \cite{Cheng1}.

In this paper, we discuss another possibility of solving the
problem of the lifetime ratio of ${\sl \Lambda}_b$ to $B_d$.
In the literatures \cite{BBSUV1}--\cite{Colan1}, the expansion parameter 
$\Delta$ in eq.(\ref{eqn:gener}) is taken as $m_b$. 
However, the physical reason of the setting $\Delta = m_b$ is not so obvious. 
Therefore the value of $\Delta$ should be determined such that it is
able to reproduce the experimental results of all lifetime ratios of
bottomed hadrons decaying only through the weak interactions. 
For bottomed hadrons,
the other lifetime ratios which have ever been measured by 
experiment \cite{LEP1} are
\begin{eqnarray}
  \frac{\tau\left( B^- \right)}{\tau \left( B_d \right)} &=& 
   1.09 \pm 0.02  \,,  \label{eqn:ratiobbexp} \\
  \frac{\tau\left( B_s \right)}{\tau \left( B_d \right)} &=& 
   0.97 \pm 0.05 \,.  \label{eqn:ratiobbsexp}
\end{eqnarray}
If the $B_i$, $\varepsilon_i$ and $r_{{\sl \Lambda}_b}$ of 
the $1/\Delta^3$ order terms are treated as completely free parameters,
we can not obtain any restriction to the value of $\Delta$.
However the $B_i-1$ and $\varepsilon_i$ shows  
how large non-factorizable effects contribute to the meson matrix elements
of local four-quark operators, since if the factorization hypothesis 
is valid, $B_i = 1$ and $\varepsilon_i = 0 $. 
Therefore these parameters should be given some physical restriction.
In order to restrict the value of these parameters, we use 
the estimations by QCD sum rules\cite{Colan1,BLLS1} and 
non-relativistic quark model\cite{Rosner1,Cheng1}. 
For $B_i$ and $\varepsilon_i$, the estimations by QCD sum 
rules\cite{BLLS1} 
\begin{eqnarray}
   | B_i -1 | &\sim& 10^{-2} \,, \nonumber \\
   | \varepsilon _i | &\sim& 10^{-2} \,,
\end{eqnarray}  
indicate that non-factorizable contributions to the meson matrix elements
of local four-quark operators are not so large. Therefore we give 
restrictions to these parameters as follows,
\begin{eqnarray}
   | B_i -1 | &\leq& 0.1 \,, \nonumber \\
   | \varepsilon _1 | &\leq& 0.1 \,, \nonumber \\
   | \varepsilon _2 | &\leq& 0.05 \,. \label{eqn:paramrest1}
\end{eqnarray}  
For the parameter $r_{{\sl \Lambda}_b}$, $r_{{\sl \Lambda}_b} \sim 0.1 - 0.3$
has been obtained by using QCD sum rules\cite{Colan1} and 
$r_{{\sl \Lambda}_b} \sim 0.6$ by using non-relativistic quark 
model\cite{Cheng1}. Therefore we give restriction to the parameter 
as follows,
\begin{equation}
  0.1 \leq r_{{\sl \Lambda}_b} \leq 0.6 \,. \label{eqn:paramrest2} 
\end{equation}
Under setting the regions (\ref{eqn:paramrest1}), (\ref{eqn:paramrest2})
to parameters $B_i$, $\varepsilon_i$, and $r_{{\sl \Lambda}_b}$  
and setting $\widetilde B_{{\sl \Lambda}_b}$ equal to 1, 
we calculate the OPE expansion parameter $\Delta$ to satisfy the 
experimental results of the lifetime ratios of $B^-$, $B_s$ 
and ${\sl \Lambda}_b$ to $B_d$. As the result, we obtain 
\begin{equation}
  \Delta = 3.25 \pm 0.67 \,\mbox{GeV} \,. \label{eqn:delta}
\end{equation}
This shows that the parameter $\Delta$ is smaller than 
the pole mass of bottom quark 
$m_b = 4.8 \pm 0.2 \,\mbox{GeV}$\cite{NS1}.

Let us consider the physical meanings of $\Delta$. 
Blok and Shifman\cite{BS1} have discussed the role of the expansion 
parameter $\Delta$ to study the effects of the subleading operators
in the inclusive heavy hadron decays. In Ref.\cite{BS1}, the parameter
$\Delta$ has been taken as
\begin{equation}
   \Delta = m_b - m_c
\end{equation}
and the analyses have been carried out at large $m_b$ limit
\begin{equation}
  \Delta = m_b - m_c \sim m_b \gg \Lambda_{QCD} \, .
\end{equation}
This limit corresponds to neglect the mass ratio $m_c/m_b$.
However this ratio is preserved under taking the heavy quark 
limit $m_{b,c} \rightarrow \infty $. 
Therefore we can not neglect it when considering
the higher order corrections in $1/m_{Q}$ expansion.     
In fact, the pole mass of charm quark $m_c = 1.4 \,\mbox{GeV}$ \cite{NS1} 
is not so small compared to $m_b$. 
The difference between quark pole masses 
$m_b - m_c = 3.40 \pm 0.06 \,\mbox{GeV}$ \cite{NS1} 
well corresponds to the $\Delta$ given in eq.(\ref{eqn:delta}).
This result indicates that we can not neglect charm quark mass 
for calculating the lifetime ratios of bottomed hadrons. 
\vspace{1cm}


If the lifetime difference of $B_d$ and ${\sl \Lambda}_b$ can be explained on 
the basis of the present approach, its applicability to other heavy hadrons 
should be investigated.
Here we take the baryon ${\sl \Omega}_b$ as a good candidate to implement 
our purpose. The baryon ${\sl \Xi}_b $ is also a candidate which decays only 
through the weak interaction. For ${\sl \Xi}_b$, however, we have to solve 
the mixing problem between ${\sl \Xi}_b$ and ${\sl \Xi}_b'$ \cite{IM1}. 
Thus it is difficult to expect that we obtain the meaningful
result for ${\sl \Xi}_b$.

The ratio $\tau({\sl \Omega}_b)/\tau(B_d)$ will give 
the important clue to the lifetime problem in $B_d$-meson 
and ${\sl \Lambda}_b$-baryon.
By using the model independent parameterization similar to 
eq.(\ref{eqn:ratiobl}), this lifetime ratio can be written in the form,
\begin{eqnarray}
   \frac{\tau \left({\sl \Omega}_b\right)}{\tau \left(B_d\right)} &=& 1
   + \frac{\mu_\pi^2 \left({\sl \Omega}_b\right) 
   - \mu_\pi^2 \left(B_d\right)}{4 m_b^2} 
   + \left(\frac{1}{4} + c_M
   \frac{m_b^2}{\Delta^2}\right)
   \frac{\mu_G^2 \left(B_d\right) 
   - \mu_G^2 \left({\sl \Omega}_b\right)}{m_b^2} \nonumber \\
   &&+ \frac{1}{\Delta^3}\left\{k_1 B_1 + k_2 B_2 + k_3 \varepsilon_1 
   + k_4 \varepsilon_2  + \left( k_7+ 
     k_8 \widetilde B_{{\sl \Omega}_b} \right) r_{{\sl \Omega}_b} \right\} 
     \,,  \label{eqn:ratiobo}
\end{eqnarray}
where $c_M$ and $k_{1-4}$ are same coefficients as eq.(\ref{eqn:ratiobl}),
and $k_{7,8}$ are coefficients of matrix elements between ${\sl \Omega}_b$ 
states of the local four-quark operators. 
In order to estimate $1/m_b^2$ and $1/\Delta^2$ terms, 
we need to know the values of 
\begin{eqnarray}
      \mu_\pi^2 ({\sl \Omega}_b) - \mu_\pi^2 (B_d) , \nonumber \\
      \mu_G^2 ({\sl \Omega}_b) - \mu_G^2 (B_d)  \, . 
      \label{eqn:mpmg}
\end{eqnarray}
Since the baryon ${\sl \Omega}_b$ has not been found yet experimentally,
we take the relations
\begin{eqnarray}
   \mu_\pi^2 ({\sl \Omega}_b) &\simeq& \mu_\pi^2 ({\sl \Sigma}_b) , \nonumber \\
   \mu_G^2({\sl \Omega}_b) &\simeq& \mu_G^2({\sl \Sigma}_b) \, ,
   \label{eqn:assum}
\end{eqnarray}
which are followed by $SU(3)$ light flavor symmetry. 
 The mass formulae of the HQET are translated into the forms
\begin{eqnarray}
  \left\{\frac{1}{3}\left( 2 m_{{\sl \Sigma}_b^*} + m_{{\sl \Sigma}_b} 
  \, \right)
  - \frac{1}{3}\left( 2 m_{{\sl \Sigma}_c^*} + m_{{\sl \Sigma}_c} \, 
  \right) \right\} 
  - \left\{\frac{1}{4}\left( 3 m_{B^*} + m_{B} \, \right)
  - \frac{1}{3}\left( 2 m_{D^*} + m_{D} \,\right) \right\} \nonumber \\
  = \left\{\, \mu_\pi^2 (B_d) - \mu_\pi^2 ({\sl \Sigma}_b) \right\}
  \left( \frac{1}{2m_c} - \frac{1}{2m_b} \right) 
  + {\cal O}\left(\,  \frac{1}{m_Q^2} \,\right) \label{eqn:pibsigma}
\end{eqnarray}
and 
\begin{equation}
   \mu_G^2 ({\sl \Sigma}_b) \simeq \frac{1}{6} \left( m_{{\sl \Sigma}_b^*}^2 
   - m_{{\sl \Sigma}_b}^2 \, \right) \, . \label{eqn:sigma}
\end{equation}
 From eqs.(\ref{eqn:muvals}), (\ref{eqn:assum}) (\ref{eqn:pibsigma}) and 
 (\ref{eqn:sigma}), we obtain
\begin{eqnarray}
      \mu_\pi^2 ({\sl \Omega}_b)  - \mu_\pi^2 (B_d)  
      &\sim& 0.03 \,(\mbox{GeV})^2 , \nonumber \\
      \mu_G^2 ({\sl \Omega}_b) - \mu_G^2 (B_d) 
      &\sim& -0.25 \, (\mbox{GeV})^2\, . 
\end{eqnarray}
Here we take $m_b = 4.8 \, \mbox{GeV}$, $m_c = 1.4 \, \mbox{GeV}$ and
the values of the hadron masses given by 
Ref.\cite{DELPHI1} for $m_{{\sl \Sigma}_b}$ and $m_{{\sl \Sigma}_b^*}$ and 
Ref.\cite{Jenkins1,Falk1,PDG} for the others.
Although this estimation is
influenced by the mass difference between ${\sl \Sigma}_b$
and ${\sl \Sigma}_b^*$, this uncertainty hardly influences to the lifetime 
ratio $\tau ({\sl \Omega}_b ) / \tau (B_d)$.
For the parameters of $1/\Delta^3$ terms, we use same parameter region
for $B_i$ and $\varepsilon_i$ as the case of lifetime ratio of 
${\sl \Lambda}_b$ to $B_d$ and set $\widetilde B_{{\sl \Omega}_b} = 1$
(valence quark approximation) 
and $r_{{\sl \Omega}_b} = 0.53$ (non-relativistic quark model\cite{Cheng1}).
By substituting these values and the $\Delta$ given by eq.(\ref{eqn:delta}), 
we have the ratio 
\begin{equation}
   \frac{\tau({\sl \Omega}_b)}{\tau(B_d)} = 1.10 \pm 0.06 \,.
\end{equation}

If we take the method in Ref.\cite{Cheng1,AMPR1} 
that $m_b^5$ in front 
of eq.(\ref{eqn:gener}) is replaced with $m_{H_b}^5$ while keeping 
$\Delta = m_b$, the ratio becomes
\begin{equation}
   \frac{\tau({\sl \Omega}_b)}{\tau(B_d)} \sim \left( \frac{m_{B_d}}
   {m_{{\sl \Omega}_b}} \right)^5 \simeq 0.55 \,,
\end{equation}
where we use the mass $m_{{\sl \Omega}_b} = 6.06 \,\mbox{GeV}$
which is derived from the mass relations given by Ref.\cite{Jenkins1}.

The discrepancy between the two predictions of the lifetime ratio
$\tau({\sl \Omega}_b)/\tau(B_d)$ is quite large in contrast 
with that in the ratio $\tau({\sl \Lambda}_b)/\tau(B_d)$. 
The present approach implies 
\begin{equation}
   \tau({\sl \Omega}_b) > \tau(B_d) \, ,\label{eqn:oblb}
\end{equation}
whereas the approach of Ref.\cite{Cheng1,AMPR1} leads to 
\begin{equation}
   \tau({\sl \Omega}_b) < \tau(B_d) \,.
\end{equation}
It should be emphasized that the hierarchy of the lifetime becomes 
entirely opposite among the both approaches.
The hierarchy obtained from the conventional approach 
\cite{BBSUV1}--\cite{Colan1} is the same as the present result (\ref{eqn:oblb}).
\vspace{1cm}


In this short note, we proposed a new approach to the problem 
of the lifetime ratio  $\tau({\sl \Lambda}_b)/\tau(B_d)$. 
The main point of this approach is that the expansion parameter 
$\Delta$ of OPE is taken to be smaller than the pole mass of 
bottom quark $m_b = 4.8 \,\mbox{GeV}$, numerically 
$3.25 \pm 0.67 \,\mbox{GeV}$.
This approach could well reproduce the experimental lifetime ratio 
$\tau({\sl \Lambda}_b)/\tau(B_d) \cong 0.78$ with keeping the
lifetime ratio $\tau(B^-)/\tau(B_d) \cong 1.09$.
The large ambiguity of $\Delta$ mainly comes from the estimations
of the $1/\Delta^3$ term contributions. The operator product expansion
(\ref{eqn:gener}) is defined at scale $\mu = m_b$. Therefore when the 
expansion parameter $\Delta$ set different value from $m_b$, 
we must consider the operator rescaling effects which come from the 
renormalization group running $m_b$ to $\Delta$. We include these 
effects in our calculation. However the effects are very small 
since the difference between $m_b$ and $\Delta$ is small.

As for the lifetime ratio $\tau({\sl \Lambda}_b)/\tau(B_d)$, 
the present approach and the one of Ref.\cite{Cheng1,AMPR1} 
lead to almost the same result. To discriminate the approaches, it is 
important to measure the lifetime of ${\sl \Omega}_b$  since the prediction of 
the hierarchy of the lifetimes of  ${\sl \Lambda}_b$ and ${\sl \Omega}_b$
is opposite between these approaches. Therefore the future experiment 
of the lifetime of ${\sl \Omega}_b$ will be able to test the models clearly.
\vspace{1cm}


{\large Acknowledgement}

The authors would like to give thanks to Professors T. Matsuoka 
and M. Tanimoto for the encouraging discussions and for reading 
of the manuscript.


\end{document}